\documentclass[preprint2]{aastex631}

\usepackage[caption=false]{subfig}
\usepackage{graphicx}
\usepackage[utf8]{inputenc} 
\usepackage[T1]{fontenc}    
\usepackage{hyperref}       
\usepackage{url}            
\usepackage{booktabs}       
\usepackage{amsfonts}
\usepackage{amsmath}  
\usepackage{nicefrac}       
\usepackage{microtype}      
\usepackage{natbib}
\usepackage{placeins}
\usepackage{todonotes}

\shorttitle{CeleriteQFD for flare detection}
\shortauthors{Esquivel et al.}
\begin{document}

\title{Detecting stellar flares in photometric data using hidden Markov models }

\author[0009-0006-0387-6544]{J. Arturo Esquivel}
\affiliation{Department of Statistical Sciences, University of Toronto, Toronto, ON, Canada}
\affiliation{Data Sciences Institute, University of Toronto, Toronto, ON, Canada}

\correspondingauthor{J. Arturo Esquivel}
\email{a.esquivel@mail.utoronto.ca}

\author[0000-0003-2779-6507]{Yunyi Shen}
\affiliation{Laboratory for Information and Decision Systems, Massachusetts Institute of Technology,  Cambridge, MA, USA}

\author[0000-0001-8016-773X]{Vianey Leos-Barajas}
\affiliation{Department of Statistical Sciences, University of Toronto, Toronto, ON, Canada}
\affiliation{Data Sciences Institute, University of Toronto, Toronto, ON, Canada}
\affiliation{School of the Environment, University of Toronto, Toronto, ON, Canada}

\author[0000-0003-3734-8177]{Gwendolyn Eadie}
\affiliation{Department of Statistical Sciences, University of Toronto, Toronto, ON, Canada}
\affiliation{Data Sciences Institute, University of Toronto, Toronto, ON, Canada}
\affiliation{David A. Dunlap Department of Astronomy \& Astrophysics, University of Toronto,  Toronto, ON, Canada}

\author[0000-0003-2573-9832]{Joshua S. Speagle}
\affiliation{Department of Statistical Sciences, University of Toronto, Toronto, ON, Canada}
\affiliation{Data Sciences Institute, University of Toronto, Toronto, ON, Canada}
\affiliation{David A. Dunlap Department of Astronomy \& Astrophysics, University of Toronto,  Toronto, ON, Canada}

\author[0000-0002-1348-8063]{Radu V Craiu}
\affiliation{Department of Statistical Sciences, University of Toronto, Toronto, ON, Canada}
\affiliation{Data Sciences Institute, University of Toronto, Toronto, ON, Canada}

\author[0000-0001-8726-3134]{Amber Medina}
\affiliation{Department of Astronomy, University of Texas-Austin, Austin, TX, USA}

\author[0000-0002-0637-835X]{James R. A. Davenport}
\affiliation{Department of Astronomy, University of Washington, Box 351580, Seattle, WA 98195, USA}

\begin{abstract}
We present a \textit{hidden Markov model} (HMM) for discovering stellar flares in light curve data of stars. HMMs provide a framework to model time series data that are non-stationary; they allow for systems to be in different states at different times and consider the probabilities that describe the switching dynamics between states. In the context of stellar flares discovery, we exploit the HMM framework by allowing the light curve of a star to be in one of three states at any given time step: \textit{Quiet}, \textit{Firing}, or \textit{Decaying}. This three-state HMM formulation is designed to enable straightforward identification of stellar flares, their duration, and associated uncertainty. This is crucial for estimating the flare’s energy, and is useful for studies of stellar flare energy distributions. We combine our HMM with a \textit{celerite} model that accounts for quasi-periodic stellar oscillations. Through an injection recovery experiment, we demonstrate and evaluate the ability of our method to detect and characterize flares in stellar time series. We also show that the proposed HMM flags fainter and lower energy flares more easily than traditional sigma-clipping methods. Lastly, we visually demonstrate that simultaneously conducting \textit{detrending} and flare detection can mitigate biased estimations arising in multi-stage modelling approaches. Thus, this method paves a new way to calculating stellar flare energy. We conclude with an example application to one star observed by TESS, showing how the HMM compares with sigma-clipping when using real data. 
\end{abstract}

\keywords{TESS, M-dwarfs, Time-series, stellar flares}

\vspace{8ex}

\section{Introduction}
\label{sec:intro}

Almost all stars in the universe with convection surface produce \textit{stellar flares} --- bursts of energy emitted from the star that are thought to be caused by magnetic reconnection \citep[see, e.g.,][]{forbes1991magnetic, donati2009magnetic}. Properly estimating the energy distribution of flares as a function of a stars' mass, age, and other characteristics is fundamental for understanding (i) the evolution of stellar magnetic fields, (ii) stellar rotation and mass-loss rates, and (iii) the highly energetic radiation environment to which planets orbiting these stars are subjected. Past studies have found relationships between flare energies and decay times, frequency of flares and stellar rotation rates, and flare duration and peak luminosity \citep[e.g., see][]{Davenport2016ApJ...829...23D, Medina2020, Chang2015ApJ...814...35C}. 

Flares are detected in the time series data of a star's brightness measurements --- a sudden, sharp increase in brightness followed by a slower decay usually indicates a stellar flare. However, detecting stellar flares is complicated by the fact that most stars also exhibit small, quasi-periodic oscillations in their brightness over time. 

It is common practice to identify flares in time series data using non-parametric models, before assuming anything about their time series signature or shape \citep{Chang2015ApJ...814...35C}. Current methods to detect stellar flares in time series data rely on multi-stage data processing and ``sigma-clipping''; after the stationary and quasi-periodic part of the time series is modeled and removed (called \textit{detrending}), points lying outside a pre-defined confidence interval are highlighted as potential flares \citep[e.g.][]{Davenport2016ApJ...829...23D,Hawley2014ApJ...797..121H, Osten2012ApJ...754....4O,Walkowicz2011AJ....141...50W,Yang2017ApJ...849...36Y,Gunther2020AJ}. \cite{Medina2020} also note that in the 3-sigma approach to flare detection and flare energy estimation, the largest source of uncertainty comes from defining the end of the flare. The sigma-clipping approach may also struggle to identify compound flares, although change-point detection does not suffer from this problem \citep{Chang2015ApJ...814...35C}.

After the flares' locations in the time series are detected, a template or model for flare shape \citep[e.g.,][]{Davenport2014} is often used to estimate flare parameters. The detrending process is often done using a flexible model, such as Gaussian processes \citep[e.g. \textit{celerite}][]{Foreman-Mackey2017}. However, detrending methods are typically done before the flare detection step, and this pre-processing may absorb lower-energy flares in the time series data, and bias the energy profile. Therefore, it would be beneficial to model the trend of the time series and the flares simultaneously. 

Other methods such as change-point detection have also been explored to identify potential flares \citep{Chang2015ApJ...814...35C}. While this approach detects the most energetic flares, it can struggle to detect the medium- to low-energetic flares that are part of the flare energy distribution. The \textit{stella} software, which uses a convolutional neural network (CNN) to find flares in TESS data, is an efficient tool for finding flares, but still relies on a probability threshold for flare detection \citep{Feinstein2020AJ}.

In this work, we introduce a new approach to stellar flare detection using \textit{hidden Markov models (HMMs)}. HMMs are flexible time series models that are popular in many domains, including ecology \citep{mcclintock2020uncovering, Timo2019, leos2017multi}, health 
\citep{williams2020bayesian} and sports \citep{otting2021copula}. The advantage of using HMMs in the context of stellar flare detection is that they are more likely to detect medium- to low energy flares than traditional sigma-clipping methods. 

HMMs provide a way to model different states underlying a time series, with a probability associated to the transitions between them \citep{zucchini2017hidden}. This is a very natural scheme to approach the detection of stellar flares --- a star may be in a ``quiet'', flare-``firing'', or flare-``decaying'' state. Thus, if an HMM is fit to the light curve of a star, every point in that time series can be estimated to come from one of these three states. This allows one to discover both the firing and decaying phases of each flare. The decaying state is particularly helpful, as it helps characterize the end of a flare as the star transitions back to the quiet state. 

In addition to using an HMM, we simultaneously fit \textit{celerite} to model the quasi-periodic trend in the star's light curve. We show that this simultaneous fitting of \textit{celerite} and the HMM not only removes the need for iterative fitting when searching for flares, but also improves the \textit{celerite} fit overall. That is, \textit{celerite} does not as easily absorb small flares, nor the decaying portion of larger flares. For our entire analysis, we adopt a Bayesian approach.

This paper presents our HMM for stellar flares detection, shows its merits, and applies it to a stellar light curve measured from TESS. Our paper is organized as follows: In Section~\ref{sec:data}, we describe the data that motivated this study. In Section~\ref{sec:method}, we thoroughly describe our method; we begin with a quick overview of \textit{celerite} (Section~\ref{sec:celerite}), followed by an introduction to HMMs (Section~\ref{sec:HMMs}); then we proceed to describe our observational model (Section~\ref{sec:obsmodel}), our particular HMM (Section~\ref{sec:QFDmodel}), our model fitting techniques (Section~\ref{sec:computation}), how we identify and characterize flares (Section~\ref{sec:flares}), and the injection recovery experiment performed (Section~\ref{sec:injection}). The results of our injection recovery experiment and the application of our HMM to TESS data are presented in Section~\ref{sec:results}. We conclude with a discussion and a summary of future research directions in Sections~\ref{sec:discussion} and~\ref{sec:conclusion}.

\section{Data} \label{sec:data}

To test and demonstrate our HMM approach, we use M dwarf TIC 031381302 two-minute cadence data measured by TESS. We use this star's Pre-Search Data Conditioning Simple Aperture Photometry (PDCSAP) light curve for both our injected flare tests (Section~\ref{sec:injection}) and for a case study demonstration detecting real flares (Section~\ref{sec:case_study}). This star was chosen because it has long portions of the time series undergoing quiescent oscillations, in which simulated flares could be injected to test our method. At the same time, there are parts of the time series for this star that have known flare events --- these portions of the light curve are used to demonstrate that our HMM can recover the same flares as other methods.

\section{Methods}\label{sec:method}

To identify flares in a stellar brightness time series, we simultaneously model the quasi-periodic changes of the star and the star's flares. For the former, we use \textit{celerite} \citep{Foreman-Mackey2017} and for the latter we use the HMM described in this paper. Readers familiar with \textit{celerite} may want to skip ahead to Section~\ref{sec:HMMs}. Those readers familiar with HMMs may want to skip to our particular setup in Section~\ref{sec:QFDmodel}.

For quick reference, a list of our mathematical notation is shown in Table~\ref{tab:notation}.
 
\begin{table*}[htp]
\caption{Notation used in this paper.}
    \centering
    \begin{tabular}{ll}
    \toprule
     & Meaning \\
    \midrule
    $Y_{t}$ & observed brightness of star at time $t$\\
    $\boldsymbol{f}$ & quasi-periodic trend of the time series (modeled with \textit{celerite)}\\
    $f_t$ & \textit{celerite}-modeled trend at time $t$ \\
    $Z_{t}$ & flaring channel (time series without trend)\\
    $\boldsymbol{\mu}$ & mean flux in Quiet state \\
    $\mathcal{K}$ & kernel for \textit{celerite} \\
    $\sigma^2$ & variance of measurement noise\\
    $S_{t}$ & state of time series at time $t$\\
    $Q$ & Quiet state \\
    $F$ & Firing state \\
    $D$ & Decay state \\
    $\text{log}(\lambda)$ & log average flux increase during firing state \\
    $\text{logit}(r)$ & logit of decay rate \\
    $p_{Q|Q},p_{Q|F}$ & transition probabilities from Quiet state \\
    $p_{F|F},p_{F|D}$ & transition probabilities from Firing state \\
    $p_{D|Q},p_{D|F},p_{D|D}$ & transition probabilities from Decay state \\
    \bottomrule
    \end{tabular}
    \label{tab:notation}
\end{table*}

\subsection{Celerite and detrending}\label{sec:celerite}

To account for the star's quasi-periodic changes as well as the mean brightness, we use \textit{celerite}, a physically-motivated Gaussian Process (GP) widely used to model the trend of stellar light curves \citep{Foreman-Mackey2017}. In principle, one could use any kernel provided in  \textit{celerite}, but in this study we use the rotation kernel, the same one used in \cite{Medina2020}, which consists of a sum of two simple harmonic oscillators.

For the GP, let $\mu$, $\mathcal{K}$, and $\boldsymbol{f}$ denote the mean function, the kernel parameter, and the (latent) trend, respectively. The value of $\boldsymbol{f}$ at time $t$ is denoted $f_t$. For a single star, the observed light curve $Y_{t}$ is modeled as the sum of the trend $f_{t}$ and a flaring channel $Z_{t}$,

\begin{equation}
\label{eq:celerite}
    \begin{aligned}
    \boldsymbol{f}| \mathcal{K},\mu &\sim \text{Celerite}(\mu, \mathcal{K})\\
    Y_t &=f_t+Z_t.\\
    \end{aligned}
\end{equation}
We use priors for the parameters of $\mu$ and $\mathcal{K}$ recommended in \citet{Foreman-Mackey2017}. The HMM for the flaring channel $Z_t$ is described in Section~\ref{sec:QFDmodel}.

\subsection{An introduction to HMMs}\label{sec:HMMs}

In its basic form, an HMM is a doubly stochastic process composed of an observable, state-dependent, process $\{Z_t\}_{t=1}^T$ and a state process $\{S_t\}_{t=1}^T$. At each point in time, $t$, the time series is assumed to be in one of $N$ possible states, (i.e., the time series follows a state process $S_1, S_2, ..., S_t$). The states are taken to be discrete latent variables generated from a first-order Markov chain that evolves over time according to an $N\times N$ transition probability matrix with entries $ p_{i,j} = \text{Pr}(S_t = j| S_{t-1} = i)$, for $i,j \in \{1, \ldots, N\}$. The state at time $t=1$ is taken to be generated according to an initial state distribution. 

The observations are modeled assuming they are emitted from a set of state-dependent distributions, i.e.,
\begin{equation}
  g_n(Z_t) =  g(Z_t | S_t =n) \text{ for } n\in \{1, \ldots, N\}. 
  \label{eq:g}
\end{equation}
For example, if the HMM is a three-state model ($N=3$), then there would be three different distributions $g$, which describe a different data generating process conditioned on each state $n$. In practice, the number of states $N$ and the distributions $g$ are defined using scientific domain knowledge. The parameters of these distributions may be estimated within the model or fixed. The $N\times N$ matrix describing the probability of transition between states is also estimated in practice.

When fitting an HMM to real time series data, one can use the estimated model to obtain the state sequence $\hat s_1, \dots, \hat s_T$; the most likely, under the assumed model to underlie the observations. This is known as state decoding and can be efficiently carried out through the Viterbi algorithm (\citealt{viterbi1967error}; see also \citealt{forney1973viterbi} for a detailed description).

The latter uses the estimated state-dependent distributions $\hat g_n$ to compute the probability density $\hat g_n(z_t)$ of each observation when a specific state $n \in \{1, \ldots, N\}$ is active. The algorithm combines these probability densities with the estimated transition probabilities to recursively determine the most likely sequence leading to each possible state $n$, at each time $t \in \{1, \ldots, T\}$. This is done through recursively computing the quantities 

\begin{equation*}
    \xi_{t,n} = \left(\max_i (\xi_{t-1,i}\hat p_{i,n})\right) \hat g_n(z_t)
\end{equation*}
for $t = 2, \dots , T$, initialized with $\xi_{1,n} = \widehat{\text{Pr}}(S_1 = n)\hat g_n(z_1)$.  i.e., $\xi_{t,n}$ corresponds to the likelihood of the state sequence from time $1$ to $t-1$ most likely to lead to state $n$ being the one active at time $t$, given the observations between $1$ and $t$. The most likely full state sequence is then determined by recursively maximizing over these likelihoods, starting with 

\begin{equation*}
    \hat s_T = \underset{n = 1, \dots, N}{\arg\max} 
 \quad \xi_{T,n},
\end{equation*}

and setting 

\begin{equation*}
    \hat s_t = \underset{n = 1, \dots, N}{\arg\max} 
 \quad \xi_{t,n} \hat p_{n,\hat s_{t+1}},
\end{equation*}

for $t = T-1, T-2, \dots, 1.$

Additional references and some foundational papers about HMMs are \cite{baum1970maximization}, \cite{rabiner1989tutorial}, and \cite{zucchini2017hidden}.

\subsection{Observational Model}\label{sec:obsmodel}

In our application, we assume an observational model for the random variable $Y_t$ given state $S_t$:
\begin{equation}
    Y_t|S_t = f_t + Z_t|S_t,
    \label{eq:obs}
\end{equation}
where $f_t$ is the celerite-modeled trend at time $t$ (note that it does not depend on the underlying state), and $Z_t$ is the \textit{flaring channel}. The latter's distribution depends on whether the star is in a quiet, firing, or decaying state, and is described next.

\subsection{A Quiet-Firing-Decay HMM for Flare Events}\label{sec:QFDmodel}

We propose a three-state hidden Markov model \citep{hamilton1990analysis} for modeling flare events in the detrended light curve. Each point in the time series can result from one of three (hidden, unobservable) states: Quiet, Firing, or Decay (denoted as $Q,F$ and $D$ respectively). The $Q$ state is used to model the time series when the star is not in any flare event, while the $F$ and $D$ states are used to model the increasing and decreasing phase of a flare. 

\begin{table*}[ht]
\caption{Interpretation of transitions between states}
    \centering
    \begin{tabular}{ccccc}
    \toprule
      &  &   & to &    \\
      &  & $Q$ & $F$  & $D$  \\
      \midrule
      &$Q$ & remain quiet   & start firing    &   \textit{forbidden} \\
from  &$F$ & \textit{forbidden}  &  increased firing  &  start decaying  \\
      &$D$ & return to quiet  & start compound flare   & decaying continues   \\
      \bottomrule
    \end{tabular}
    \label{tab:transition}
\end{table*}

Recall the probabilities $p_{S_{t-1}, S_t}$ of switching between states at each step in the time series (akin to a Markov chain); they are conditional probabilities of the form  $p{S_{t-1}|S_t}$, where $S_{t-1}$ is the previous state; for example, $p_{{F}|{Q}}$ denotes the probability of transitioning to state $F$ given that the star is in state $Q$. The interpretation given to each of these transitions is illustrated in Table~\ref{tab:transition}. 

The transition from $Q$ to $F$ accounts for the firing rate of flares from the star's quiet state. Once the star is in the flaring state, the transition from $F$ to $F$ accounts for the increasing phase of a flare, while the transition from $F$ to $D$ accounts for the decay of the flare. When the time series is in the decay state,transiting from $D$ to $F$ corresponds to a compound flare. Note that we forbid the transition from $Q$ to $D$ (i.e., when the star is quiet, it will not suddenly ``decay'') and from $F$ to $Q$ (i.e., when the star is flaring, the flare will not spontaneously disappear). All other transitions are allowed, and are each modeled with parameters that account for different physical characteristics.

We model the flaring channel $Z_t$ at time $t$ given the states $S_t\in \{Q,F,D\}$ and the previous step $Z_{t-1}$ as follows:

\begin{equation}
\label{eq:states}
    \begin{aligned}
    Z_t|(S_t=Q,Z_{t-1}, \sigma^2) ~&\sim \mathcal{N}(0,\sigma^2),\\
    Z_t|(S_t=F,Z_{t-1},\lambda, \sigma^2) ~&\sim \mathcal{N}(Z_{t-1},\sigma^2) \; + \\& \quad \: \text{Exp}(\lambda) ,\\
   Z_t|(S_t=D,Z_{t-1},r, \sigma^2) ~&\sim \mathcal{N}(rZ_{t-1},\sigma^2).
    \end{aligned}
\end{equation}

The distributions above are the $g_n$ distributions mentioned in Section~\ref{sec:HMMs} and eq.~(\ref{eq:g}). Note that the value of the time series at time $t$ is always dependent on the current state of the star $S_{t}$ and on the value of the time series in the previous step ($Z_{t-1}$) when in states $F$ or $D$.  The $Q$ state is modeled as an independent normal random variable with variance $\sigma^2$. When formulating the distributions in (\ref{eq:states}) we assume that: when quiet, the flaring channel will just be measurement noise on top of the quiescent trend; when firing, the flaring channel will be around the previous channel plus an independent flux increase exponentially distributed; and when decaying, the flaring channel will be centered around a scaled value of the previous channel.

\begin{figure}
    \centering
    \includegraphics[scale=0.29]{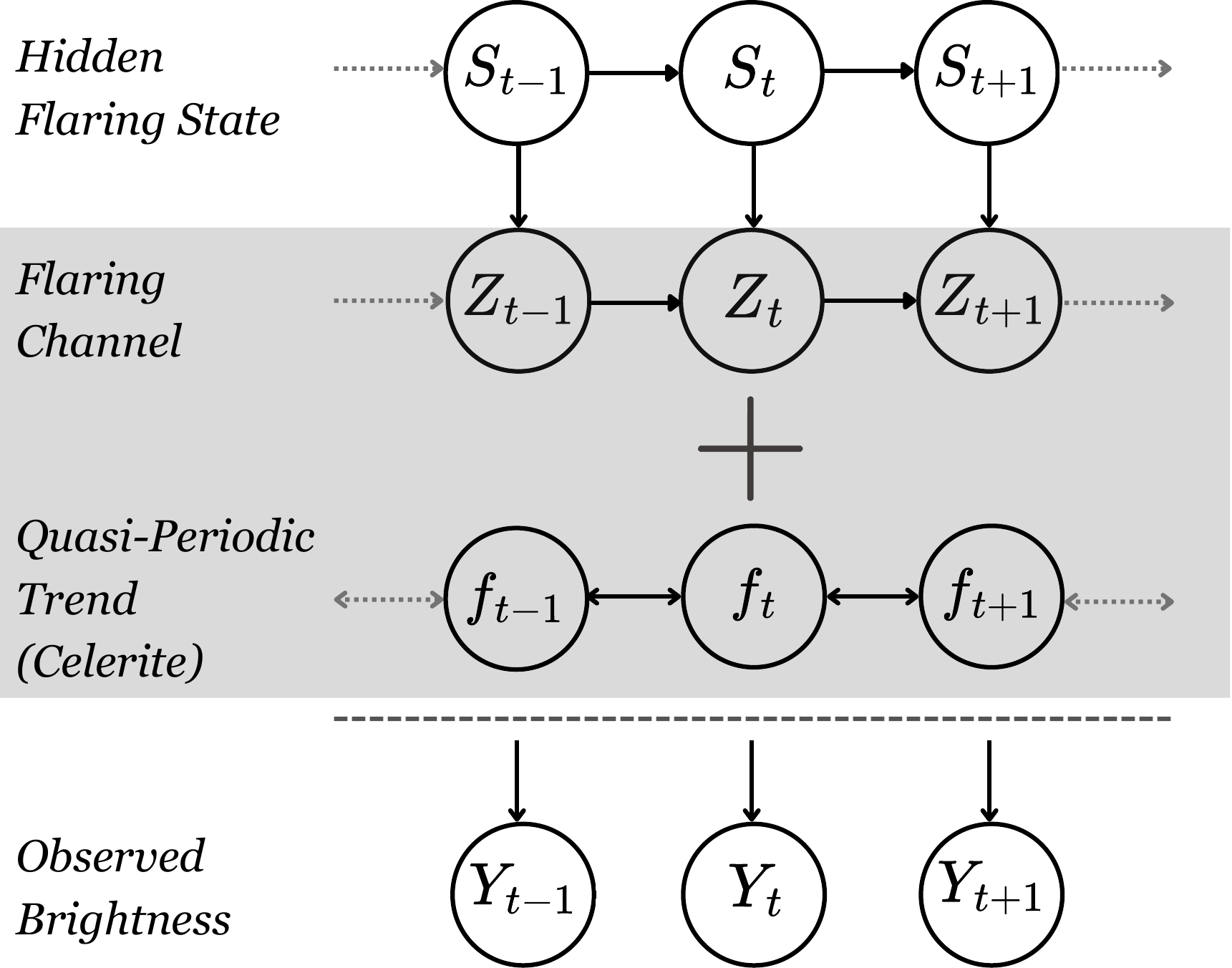}
    \caption{Graphical representation of our complete model for observed brightness decomposed into its various parts. The Quasi-Periodic Trend reflects the average brightness of the star when it is not flaring. The Flaring Channel represents the extra brightness due to the state of the star. The Hidden Flaring State represents the (unobserved) state of the star (Q,F, or D).}
    \label{fig:graph}
\end{figure}

Recall that the model for the observed light curve $y_t$ is the combination of the trend and the flaring channel; that is, \textit{celerite} and the HMM are fit simultaneously, such that $\boldsymbol{f}$ is a latent variable (Figure~\ref{fig:graph}). We call this combined model \textit{celeriteQFD}. A summary of the HMM parameters, their prior distributions, and hyperparameter values can be found in Table \ref{tab:prior}. $\text{invGamma}$ is the inverse Gamma distribution and the transition probabilities (e.g., $p_{Q|F},p_{F|D}$, etc.) use Dirichlet prior distributions. This setting is used for all injection-recovery and real data examples.

\begin{table*}[htp]
\caption{Prior setting of the QFD part of the model}
    \centering
    \begin{tabular}{llll}
    \toprule
    Parameter & Meaning & Prior distribution & Hyperparameter used\\
    \midrule
    $\mu$ & Mean flux at quiet & $N(\mu_0,\sigma_0^2)$ & $\mu_0=0,\sigma_0^2=100\sigma^2$\\
    $\sigma^2$ & variance of measurement noise & $\text{invGamma}(\alpha,\beta)$ & $\alpha=.01,\beta=.01$\\
    $\log(\lambda)$ & log average flux increasing during firing & $N(\mu_\lambda, \sigma^2_\lambda)$ & $\mu_\lambda=0,\sigma^2_{\lambda}=1e3$\\
    $\text{logit}(r)$ & logit of decay rate & $N(\mu_r,\sigma_r^2)$ & $\mu_r=0,\sigma^2_{r}=1e3$\\
    $p_{Q|Q},p_{Q|F}$ & transition probabilities from Quiet state & $\text{Dir}(\alpha_Q)$ & $\alpha_Q=(1,0.1)$\\
    $p_{F|F},p_{F|D}$ & transition probabilities from Firing state & $\text{Dir}(\alpha_F)$ & $\alpha_F=(1,1)$\\
    $p_{D|Q},p_{D|F},p_{D|D}$ & transition probabilities from Decay state & $\text{Dir}(\alpha_D)$ & $\alpha_D=(1,0.1,1)$\\
    \bottomrule
    \end{tabular}
    \label{tab:prior}
\end{table*}

\subsection{Computation, Model Fitting, and State Decoding}\label{sec:computation}

For computational purposes, we split each time series $\{y_t\}_{t=1}^T$ into smaller chunks of 2000 time steps. We subtract the overall mean from the light curve to center it around zero. Our model was implemented in \texttt{stan} \citep{carpenter2017stan}, while the C++ code for \textit{celerite} was adopted from the Python library \texttt{EXO-PLANET}. The posterior distributions of model parameters and derived quantities are sampled using a dynamic Hamiltonian Monte Carlo algorithm \citep{hoffmannuts}. States are decoded using the Viterbi algorithm as described in Section~\ref{sec:HMMs} \citep[see also][]{forney1973viterbi}. The \texttt{stan} implementation as well as the injection recovery tests can be found in the first author's github repository \href{https://github.com/Esquivel-Arturo/celeriteQFD}{Esquivel-Arturo/celeriteQFD}. 

\begin{figure*}[!htb]
    \centering
    \includegraphics[scale=0.7]{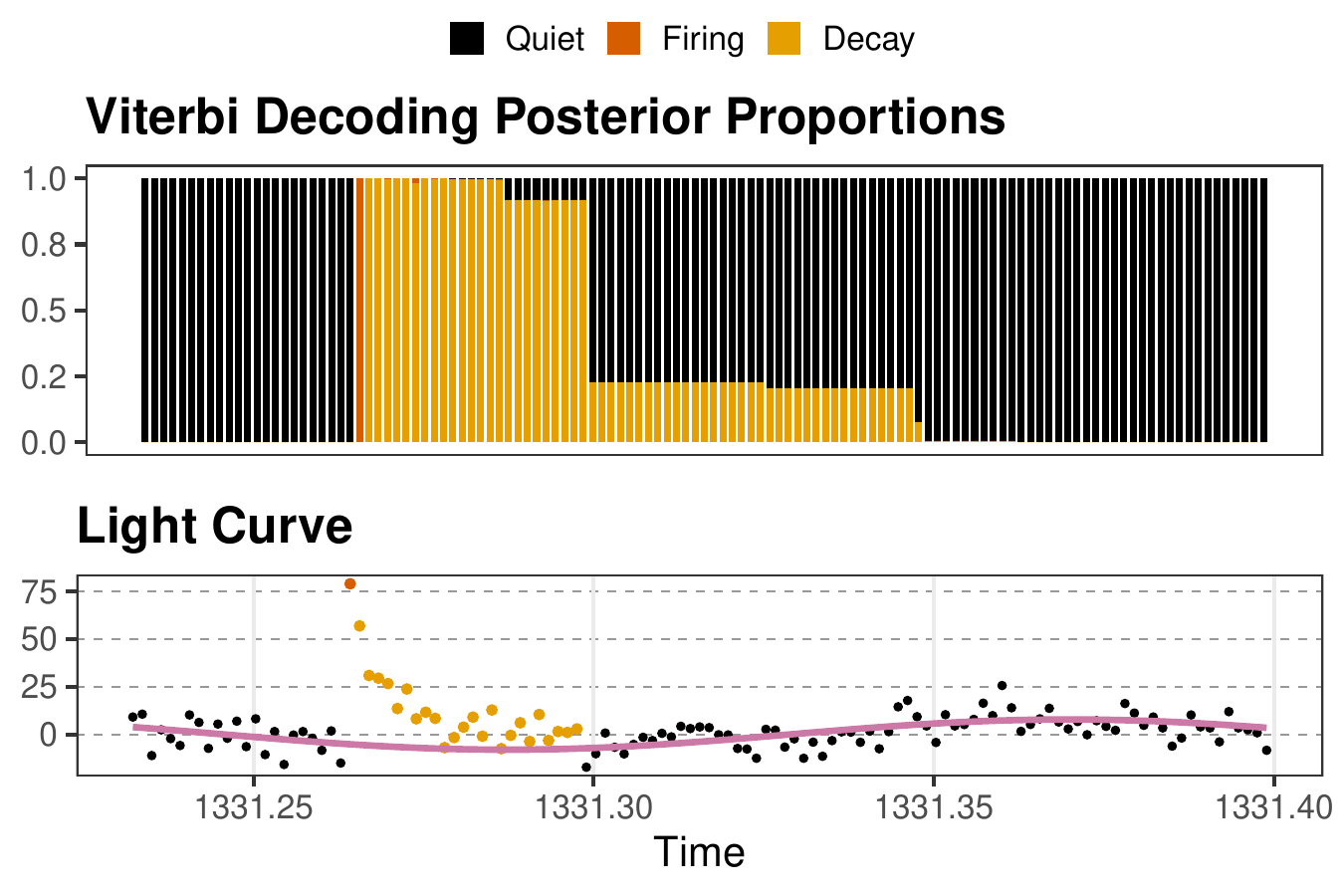}
    \caption{Detected flare example from implementing \textit{celeriteQFD} on the real time series of TIC 031381302 \textit{Top panel:} State "decoding distributions" across all Viterbi sequences per observation. \textit{Second panel:} shows the fit of \textit{celeriteQFD}, the estimated trend (purple curve) and assigned state to each point (black points are state $Q$, red points $F$, and orange points are $D$).}
    \label{fig:flare_uncer_large}
\end{figure*}

To obtain samples from the posterior distributions of the model parameters, we use two MCMC chains and sample 1000 (2000 in total) posterior samples, after discarding the first 1000 samples obtained during the warm-up period. For each joint posterior sample of the parameters, we use the Viterbi algorithm to uncover the most likely state sequence that could have generated the data. In this manner we are able to propagate the uncertainty around our parameters to produce 2000 most likely state sequences and capture the variability in state decoding results. Thus, for every point in the time series we have a "decoding distribution" (see Figure~\ref{fig:flare_uncer_large} top panel)  of the states ($Q$,$F$,$D$). For each point in the time series, we estimate the state of the star to be the one that appears most frequently across the 2000 decodings of that time step.  

Each \textit{celeriteQFD} implementation on 2000 time steps took between 1 and 4 hours to run using two cores of a M1 MacBook Pro with 16 GB of memory. 

\subsection{Identifying and Characterizing Flares}\label{sec:flares}

To identify flares after fitting \textit{celeriteQFD}, we consider, all consecutive points decoded in a non-$Q$ state (i.e., the rise $F$ and fall $D$ of the flare) are used to define the duration of a flare. A flare is considered over once the time series returns to state $Q$. In other words, the duration of a flare is defined as the time elapsed from when the star enters the flaring state ($Q$ to $F$) to when the star re-enters the quiet state ($D$ to $Q$). For example, in Figure~\ref{fig:flare_uncer_large} the flare was estimated to commence with the peak red point and end right before time 1331.30. 

\begin{figure*}[!htb]
    \centering
    \includegraphics[scale=0.7]{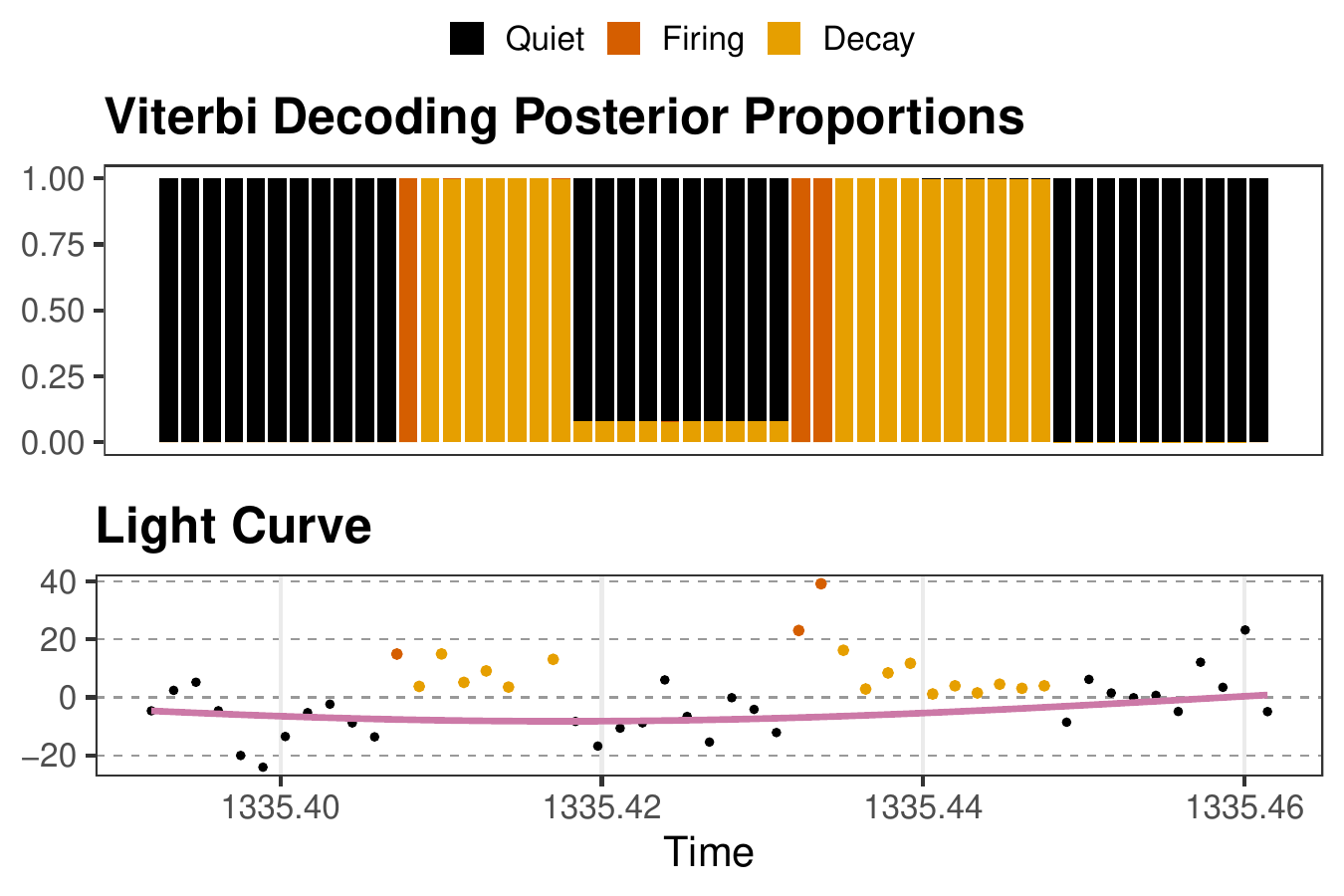}
    \caption{Example of two flares detected from implementing \textit{celeriteQFD} on the real time series of TIC 031381302. \textit{Top panel:} State "decoding distributions" across all Viterbi sequences per observation. \textit{Second panel:} shows the fit of \textit{celeriteQFD}, the estimated trend (purple curve) and assigned state to each point (black points are state $Q$, red points $F$, and orange points are $D$).}
    \label{fig:flare_uncer2}
\end{figure*}

This method also allows us to find compound flares (e.g., the time series could be in the decaying phase of a flare, and then start firing again). As the state decoding of a compound flare will have multiple peaks we skip any peaks identified in a flare's duration when searching for the next flare. We can also quantify uncertainty around the Viterbi "decoding distributions", e.g., Figure~\ref{fig:flare_uncer2} shows two flares detected very close to each other. From the decoding proportions we can see 10\% of the Viterbi state decodings point to a compound flare instead of two separate flares. 

\subsection{Injection Recovery Experiment}\label{sec:injection}

To test the ability of our HMM to detect stellar flares, we perform an injection recovery experiment: we inject simulated flares into a real stellar time series, apply our HMM algorithm to detect the simulated flares, and compare our results to the ground truth and those of using a sigma-clipping approach. 

We use the mean-centered flux time series data from one star as the base time series for our injection recovery experiment, and randomly inject Kepler flares following the procedure outlined in \cite{Davenport2014}. We use TIC 031381302 (from day 1325.292 to day 1327.377, $n=1501$). This time segment was chosen to avoid already-discovered natural flares. 

We inject five flares, both small and large, at randomly chosen points in the time series. The time scale of our Kepler-like flares, $t_{1/2}$, is proportional to its peak flux. The peak fluxes of the injections are \textit{i.i.d.} Pareto (i.e., they follow a truncated power-law distribution):
\begin{equation}
\label{eqn:pareto}
    p(x)=\frac{\alpha x_m^\alpha}{x^{\alpha+1}},
\end{equation}
where $x_m$ and $\alpha$ are the distribution parameters. 

We use different parameter values for $(t_{1/2}, x_{m},\alpha, \delta, \beta)$ to simulate small and large flares. We use $(5\times10^{-5},10,1, 30, 150)$ for small flares, and $(5\times10^{-5},50,1, 0, 300)$ for large flares.

For each set of parameter values, we separately simulate and inject five flares into the time series, and perform our analysis of flare recovery. We repeat the procedure 100 times for each parameter scheme. Although we do not explicitly study our method's ability to recover compound flares, we do allow the simulated flares to overlap in time and form compound flares in the base time series.

\subsection{Flare Detection Evaluation}

For each simulated time series, we run \textit{celeriteQFD} and obtain a Bayesian estimate of the state of the star ($Q$,$F$, or $D$) at every time step (see Section~\ref{sec:computation} and Figure~\ref{fig:flare_uncer_large}). Once we have the estimated states for all points in a time series, flares are identified as described in Section~\ref{sec:flares}.

To evaluate the accuracy of flare detection using our HMM framework (Section~\ref{sec:HMMs}), we compare the number of detected flares to the ground truth (i.e., flares injected). For each fitted model (100 per method and flare scheme), we calculate the true positive, false positive, and false negative rates of detection, as well as the sensitivity and the positive predictive value (PPV):

\begin{eqnarray}
    sensitivity &=& \frac{TFD}{TFT}, \label{eq:sens} \\[10pt] 
    PPV &=& \frac{TFD}{FD},
    \label{eq:PPV}
\end{eqnarray}
where $TFD$ is the number of true flares detected; $TFT$ the total number of true flares; and $FD$ the total number of flares detected. We also assess flare detection in terms of the full duration of the flaring processes, i.e. we compute the per observation sensitivity and PPV:  

\begin{eqnarray}
    sensitivity &=& \frac{TFD_o}{TFT_o}, \label{eq:sens2} \\[10pt]
    PPV &=& \frac{TFD_o}{FD_o},
    \label{eq:PPV2}
\end{eqnarray}
where $TFD_o$ is the number of observations part of a true flare correctly identified; $TFT_o$ is the total number of observations that are part of a true flare; and $FD_o$ is the total number of observations identified to be part of a detected flare. Note that sensitivity and PPV ((\ref{eq:sens}) and (\ref{eq:PPV})) should each be 1 if we perfectly identify all true flares. Similarly, the per observation metrics ((\ref{eq:sens2}) and (\ref{eq:PPV2})) should be 1 if the entire duration of every flare is correctly identified. 

We compare our results to those obtained with the more commonplace sigma-clipping $a$--$b \sigma$ rule, where $a$ is the number of consecutive points in the time series that are $b \sigma$ away from the mean flux $\mu$ of the detrended time series (i.e., the rule outlined in \citealt{Chang2015ApJ...814...35C} and used in e.g. \citealt{Medina2020,ilin2019flares}). In particular, we compare against using a $1$--$3 \sigma$ approach and when fitting \textit{celerite} for detrending in the this method, we use the same priors and kernel as with \textit{celeriteQFD}.

\section{Results}
\label{sec:results}

\subsection{Injection recovery}

\begin{figure*}[!htb]
    \centering
    \includegraphics[scale=0.6]{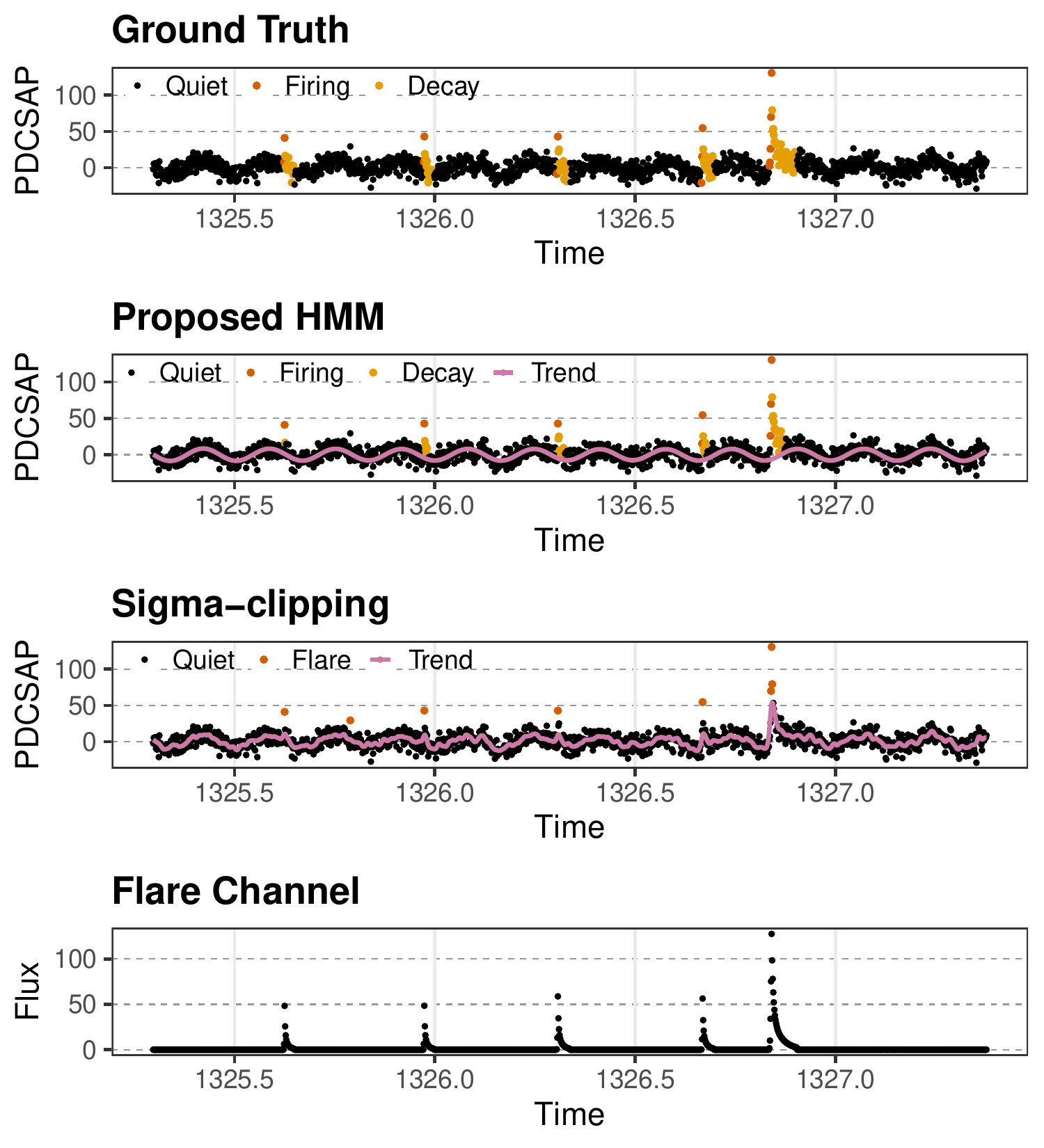}
    \caption{An example of one injection recovery simulation in the small flare scheme. \textit{Top panel:} ground truth --- the real time series for TIC 031381302 with five simulated flares injected (their real states colour-coded). \textit{Second panel:} shows the fit of our proposed algorithm that simultaneously models the trend with \textit{celerite} (purple curve) and assigns states to each point in the time series (black points are state $Q$, red points $F$, and orange points are $D$). \textit{Third panel:} the sigma-clipping approach that uses \textit{celerite} alone to model the trend (purple curve), with outliers beyond $3\sigma$ (red points) used to identify flares. \textit{Bottom panel:} the flare channel that was injected into the time series.}
    \label{fig:inj_small_eg}
\end{figure*}

 An example of one injection recovery experiment under the small flare scheme is shown in Figure~\ref{fig:inj_small_eg}. The flux was centered, i.e. it is the raw flux data from which we subtracted the grand mean. It shows the estimated trend and state sequence from our HMM (second panel), along with the ground truth (top panel) and with the result of using a $1-3\sigma$ approach (third panel). We use $1-3\sigma$ because it is more sensitive than the usual $3-3\sigma$ and so it is more likely for it to detect small injections. In this example it can be seen how, through state $D$, the HMM directly identifies a larger proportion of flaring events than sigma-clipping. Moreover, note that \textit{celerite} can absorb part of the flares into the estimated trend of the time series, and so reduce the chance that a flare is detected by sigma-clipping rules.

\begin{figure*}
\centering 
\subfloat[Per flare]{%
  \includegraphics[width=1\columnwidth]{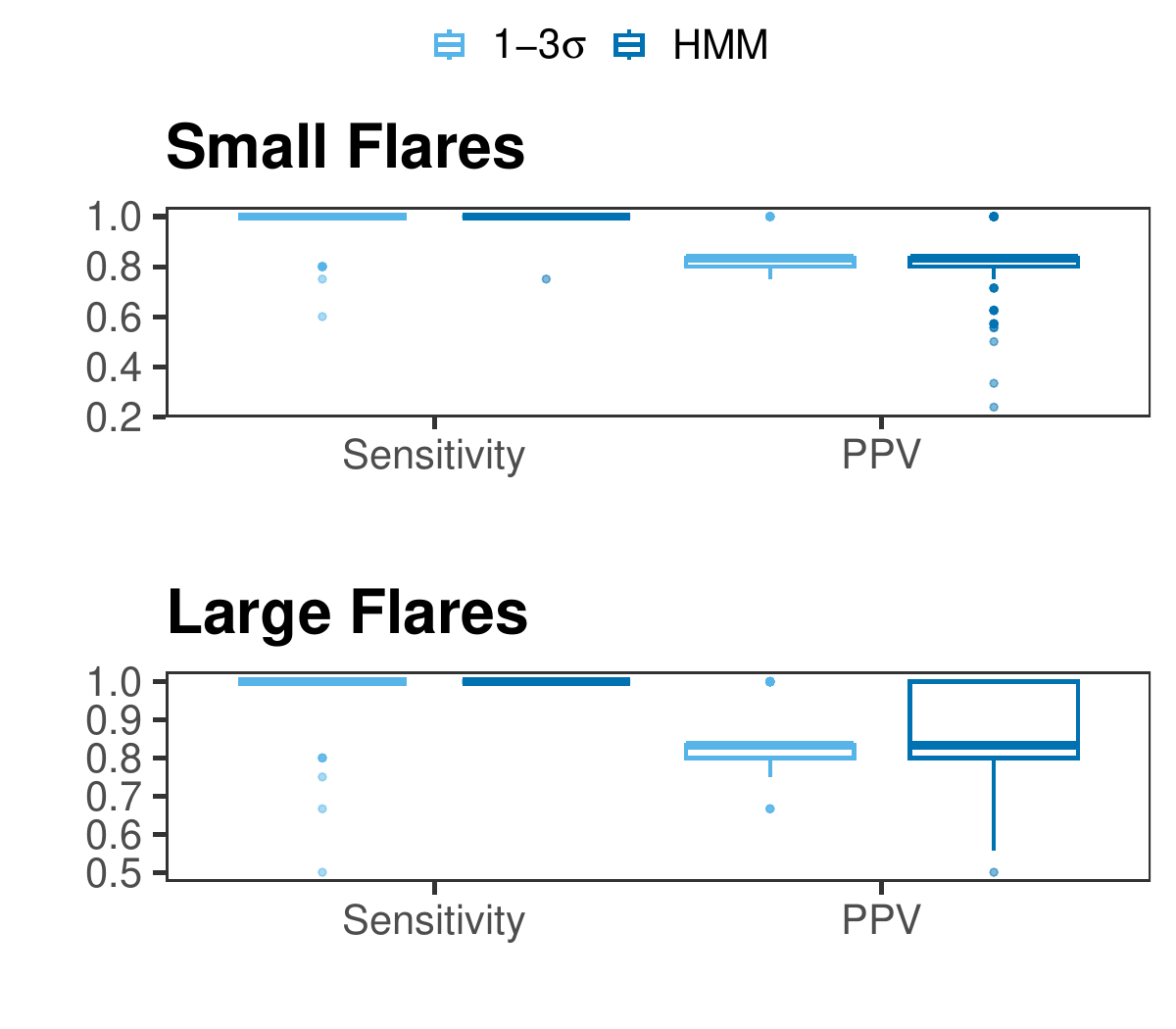}%
  \label{plot:sval}%
}\qquad
\subfloat[Per observation]{%
  \includegraphics[width=1\columnwidth]{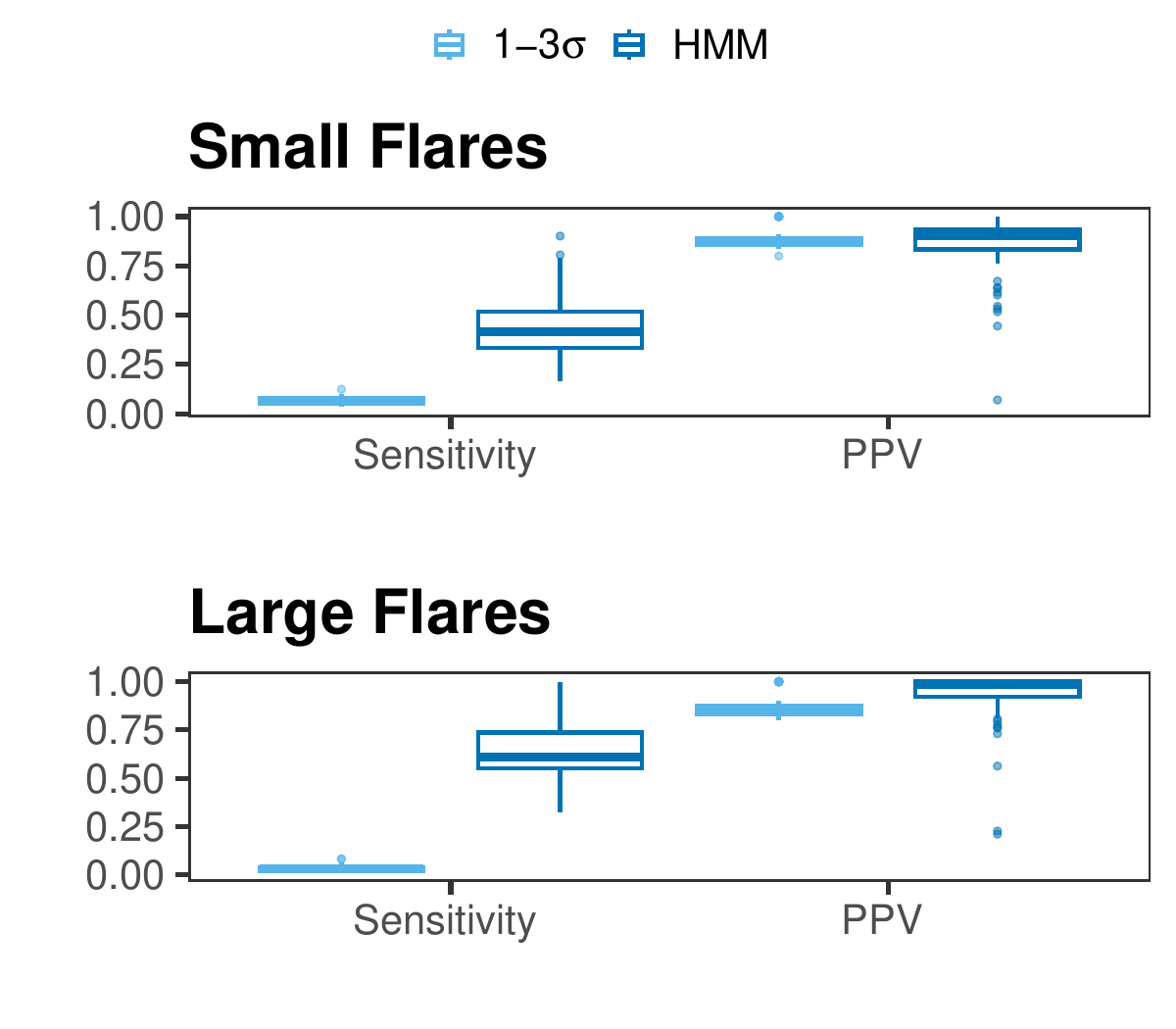}%
}
\caption{Flare recovery sensitivity and positive predictive value (PPV) distributions across 100  small and large flares injections, using the 1-3$\sigma$ rule and \textit{celeriteQFD} (HMM). \textit{(a)} shows results from detecting flare occurrences (equations (\ref{eq:sens}) and (\ref{eq:PPV}) are used). \textit{(b)} shows results on a per-observation basis (equations (\ref{eq:sens2}) and (\ref{eq:PPV2}) are used).}
     \label{fig:res_inj_rec_loc}
\end{figure*}

The results of the injection recovery experiments are presented in Figure~\ref{fig:res_inj_rec_loc}. The box plots show a comparison of our method's detection performance with that of $1-3\sigma$ clipping for small and large flares schemes. They show the sensitivity and PPV distributions across the 100 simulations of each setting. e.g., the first (light-blue) box on the top-left panel corresponds to the distribution of the per flare sensitivity (see (\ref{eq:sens})) across all 100 sigma-clipping models fitted to the 100 injections of five small flares simulated. 

From part (a) of Figure~\ref{fig:res_inj_rec_loc}, it can be seen that both methods perform almost ideally when it comes to detecting the occurrence of large and small flaring events. Flare detection sensitivity, i.e., the probability of correctly detecting a flare was computed to be 1 in almost all the 100 simulations. They also achieve a similar performance in terms of the PPV (see (\ref{eq:PPV})); i.e., the probability of an identified flare occurrence being correct. However, under the large flares scheme, \textit{celeriteQFD} often performed slightly better, achieving PPVs of 1 (it was 0.8 almost always with sigma-clipping). This indicates that our method is less susceptible to producing false detections. 

These results suggest our proposed HMM framework is at least as good as sigma-clipping for flaring events detection tasks. But the meaningful difference of our method consists of its ability to directly provide an estimate of the full duration of flaring events. Figure~\ref{fig:res_inj_rec_loc} part (b) contains the distributions of the per observation performance metrics (see (\ref{eq:sens2}) and (\ref{eq:PPV2})). i.e., the metrics are computed using all points identified as part of a flare by the methods, as compared with all the points that truly belong to injected flares. The plots show that both methods very rarely flagged observations outside of a real flare (PPVs are concentrated very close to 1). Also one can clearly see the difference between methods when it comes to spotting all light points that are part of a flare; \textit{celeriteQFD} consistently identified more than 50\% of the observations forming part of a small flare and close to 70\% for large flares. 

Sigma-clipping alone is never used for a full characterization of a flare, which is usually done through further data modeling steps (see \cite{Chang2015ApJ...814...35C} for example). Still, this experiment demonstrates the capacity our method has to describe the entire duration of detected flares without the need for extra steps.

Another crucial difference of our method is that it simultaneously carries out detrending and flare detection. By considering the light curve observations to be a combination of the long-term trend and a flaring channel (see (\ref{eq:obs})), both components are modeled accounting for the effects of the other part on the observational process. i.e., the posterior distributions of the \textit{celerite} parameters contain information on the HMM parameters and vice-versa. 

\begin{figure*}
    \centering
    \includegraphics[scale=0.85]{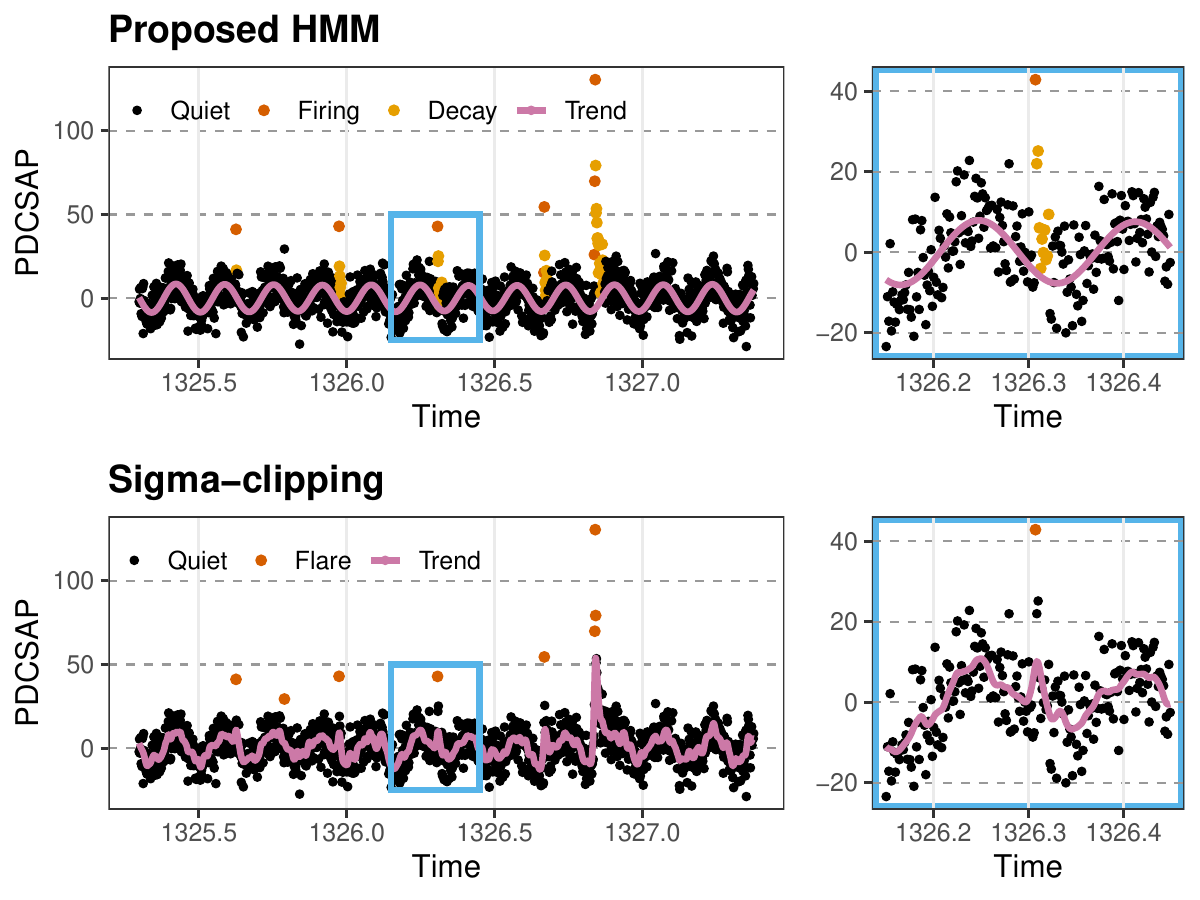}
    \caption{An example comparison of \textit{celeriteQFD} and standard sigma-clipping for identifying the injected flares to part of the TIC 031381302 light curve. \textit{Top row:} \textit{celeriteQFD}, that simultaneously models the trend with \textit{celerite} and assigns states to each point in the time series. \textit{Bottom row:} the sigma-clipping approach that uses \textit{celerite} alone to model the trend, with outliers beyond $3\sigma$ used to identify flares. The right-hand column shows a zoomed-in portion of one of the flares identified in both methods.}
    \label{fig:sim_flare_detection_close}
\end{figure*}

Figure~\ref{fig:sim_flare_detection_close} illustrates this by showing a zoomed-in flare detected by both methods from the example in Figure~\ref{fig:inj_small_eg}. The first thing to note, is that, overall, the \textit{celerite-}estimated trend is considerably less affected by light curve variability when fitted simultaneously with the HMM. Moreover, in the zoomed flare, note that \textit{celerite}, when used alone, can absorb part of the flares into the trend. Since the data points correspond to an injected flare, we know the increased flux of those observations is not part of the long-term pattern of the light curve and that detrending should ideally ignore it. \textit{celeriteQFD}, is not only able to estimate the trend unaffected by the increased flux, it is also capable of identifying many of the observations as part of the decaying phase of a flare. This result is critical, suggesting that better detrending can be achieved, further leading to less biased estimates of the flares' energies.

\subsection{Case Study: Photometric Data from the TESS Mission}
\label{sec:case_study}
As a demonstration of real flares detection, we apply our HMM method to a large portion of the TIC 031381302 light curve. The time series was sliced into pieces of 2000 time steps to more efficiently fit the models and conduct flare detection. Trace plots of the MCMC sub-chains corresponding to the parameters were produced and inspected without any indication of lack of convergence. 

\begin{figure*}[!htb]
    \centering
    \includegraphics[width = .8\linewidth]{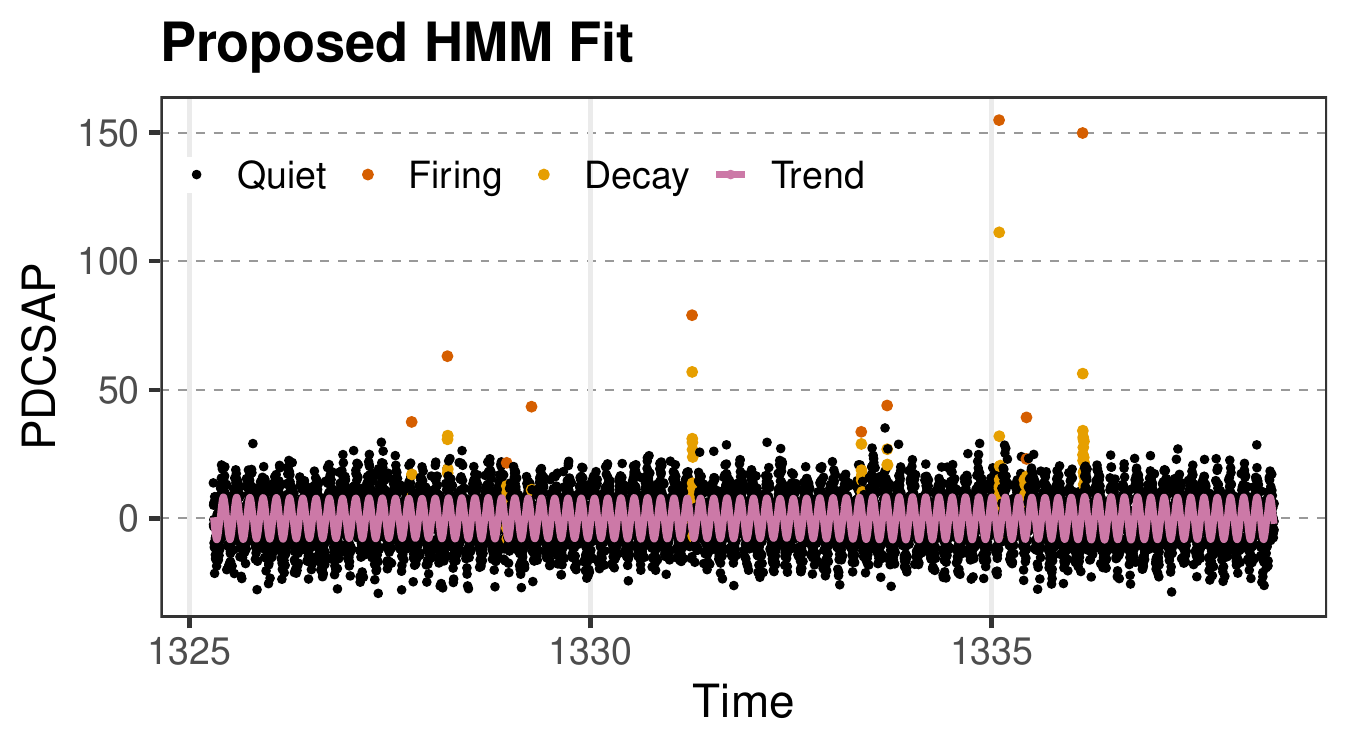}
    \caption{TIC 031381302 mean centered light curve along with the fit of \textit{celeriteQFD} that simultaneously models the trend with \textit{celerite} and assigns states to each point in the time series.}
    \label{fig:flare_detection}
\end{figure*}

A large portion (from day 1325.292 to day 1353.177) of TIC 031381302 PDCSAP mean-centered flux, along with the resulting fit and state decoding, are shown in Figure~\ref{fig:flare_detection}. During the period observed, a total of 11 flares were detected, with an average duration of 10.3 observations (approximately 0.01545 days or 1334.88 seconds). The estimated (using the posterior median) transition probability ($p{Q|F}$) was $0.00170$, with a 95\% credible interval of $(0.00007, 0.00820)$. i.e., the estimated probability that this star starts firing at any particular time, given that it was quiet in the previous time step, is 0.17\%.

\begin{figure*}[!htb]
    \centering
    \includegraphics[scale=0.85]{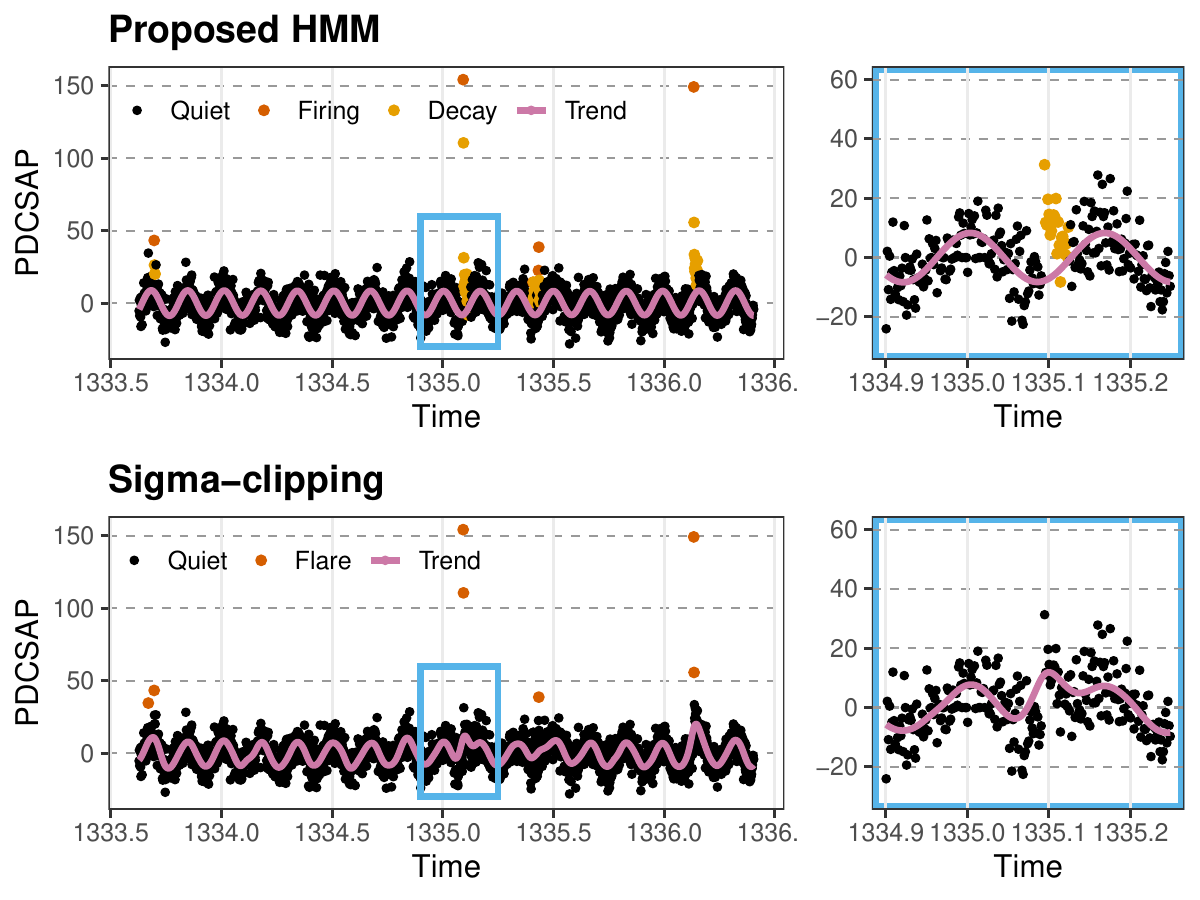}
    \caption{An example comparison of our HMM approach and standard sigma-clipping for identifying flares, using real data part of the TIC 031381302 light curve. \textit{Top row:} \textit{celeriteQFD}, that simultaneously models the trend with \textit{celerite} and assigns states to each point in the time series. \textit{Bottom row:} the sigma-clipping approach that uses \textit{celerite} alone to model the trend, with outliers beyond $3\sigma$ used to identify flares. The right-hand column shows a zoomed-in portion of one of the flares identified in both methods.}
    \label{fig:flare_detection_close}
\end{figure*}

A smaller portion (from day 	
1333.627 to day 1336.404) of both models fitted to the TIC 031381302 are shown in Figure~\ref{fig:flare_detection_close}. The right-hand side shows a zoom-in into a portion containing what both methods identified as a flare event. Note the similarity with many aspects of the flare shown in the right-hand side of Figure~\ref{fig:sim_flare_detection_close}. The trend modeled using \textit{celerite} only (bottom row) gets distorted, absorbing observations of higher brightness. Given that this points follow two observations of peak brightness, it is rather likely that at least some of them correspond to the decaying phase of a flare. This is precisely the type of case in which flare energies could be underestimated. Also note that using our approach, the estimated trend remained unaltered in this window and the HMM identified multiple observations to be in decaying state (the same way as in the synthetic case). 

By comparing the simulation results, where the ground truth is know, with real data results, it seems \textit{celeriteQFD} can in fact better model the long-term trend of a light curve. Additionally, these results indicate the model is capable of directly identify light points conforming the decaying phase of a real flare, determining the duration of flaring events. Moreover, through the "decoding distributions" it directly provides a way to quantify the uncertainty about the estimated durations of the flares. e.g, in Figure~\ref{fig:flare_uncer_large} top panel it can be seen that almost 20\% of the Viterbi sequences estimated the decaying phase of the flare detected extends up until time 1331.35. Similarly, it can be seen that around 10\% of them identified the flare to end 9 time steps earlier than estimated using the majority state. This uncertainty can be easily propagated into the final goal of producing energy distributions, potentially leading to more reliable and comprehensive distributions.

\section{Discussion}
\label{sec:discussion}

As shown in this work, an advantage of having the HMM as a flaring model is that we can identify the whole course of a flare, via states assigned to each time point. We no longer need to cross-correlate the time series with stellar flare templates, and thus remove a step in the analysis process. The HMM approach also allows us to detect compound flares more easily. Moreover, these can be used to give a probabilistic sense to the duration of the flares, making it possible to produce more comprehensive distributions of the energies of the flares.  

The inclusion of a state associated with decaying and the capability to simultaneously perform detrending and state decoding constitute a relevant benefit of our proposed method. Through our injection recovery experiment and analysis of a real star, we have shown that \textit{celeriteQFD} can produce better and more stable estimates of the long-term trend of a light curve than \textit{celerite} alone. The agreement between the results obtained for synthetic cases (where the ground truth is known) and real data cases indicates that biased estimation of the trend is indeed an issue that can arise when detrending is done prior to flare detection. It also provides some reassurance that our model is better equipped to handle the problem and prevent such bias. Further, we have shown that our HMM method can detect flares of lower energy that might be missed by other methods, even sensitive methods such as $1-3\sigma$ clipping. Thus, the HMM approach to flare detection could be well-suited for detecting flares in more ``inactive'' G-type stars.  

It is worth mentioning that our flare recovery experiments are in no way exhaustive of flare morphologies present in stars. While it is true that stellar flare templates are more physically motivated than \textit{celeriteQFD}, and that \textit{celeriteQFD} is not a generative model, HMMs still classify states very well even when the generative process is not specified correctly \cite{ruizsuarez2022}. Also, note that we do not make any strong assumptions about the nature of flare events. The few restrictions on the matter are made through the state-dependent distributions, which parameters are estimated. This provides flexibility for the model to produce different results in different contexts. We intend to explore this further, and conduct energy recovery experiments on different types of simulated flares, assessing our method's performance and contrasting it with that of other methods. 

Considering the ultimate goal of estimating flares' energies and their distributions, this method paves a new way to calculating stellar flare energy. \textit{celeriteQFD} could prove be generally applicable and quantitatively reliable. We plan to continue this work by using our model on other light curves from different stars measured by TESS. 

\section{Conclusions \& Future Work}\label{sec:conclusion}

In this paper, we have introduced a hidden Markov model for discovering stellar flares in time series data of M-dwarf stars. Our approach has some notable advantages over previous approaches. 

First, our method simultaneously fits a \textit{celerite} model for the quiescent state of the star and a three-state flaring model through a  hidden Markov process. With this approach, we not only obtain a better estimate of the quiescient state of the star, but also eliminate the need for sigma-clipping approaches and iterative fitting of time series data. Moreover, with the combined approach of \textit{celerite} and the HMM, \textit{celerite} does not absorb early or late parts of the flares. Concurrently, the HMM better identifies the whole course of the flare, and can also identify compound flares easily.

Second, through our flare injection recovery experiment, we find that our HMM method for flare detection achieves the same or better sensitivity and positive predictive value compared to sigma clipping (Figure~\ref{fig:res_inj_rec_loc}).

Third, our method enables a coherent path for uncertainty quantification. Rather than providing a single most-likely flare state sequence for the time series, we obtain a posterior distribution of most-likely flare state sequences by propagating the uncertainty from our parameter estimates. This allows us to capture the variability and uncertainty on the duration of each flare.

While our approach has significant advantages, one potential disadvantage is computation time. Currently, it can take on the order of a couple of hours to run the HMM model on a single star's time series as measured by TESS, on modest resources (see Section~\ref{sec:computation}). We are currently exploring approaches to overcome this challenge, as the advantages of our method seem to outweigh this minor (and surmountable) drawback.

Another potential criticism of our approach is that we have not developed the HMM to realistically simulate stellar time series with flares. That is, in this work, we are measuring the capacity of our proposed HMM plus the \textit{celerite} model to identify flares well, but we are not evaluating the model's generative properties. This is something that could be improved upon and explored in the future.

Overall, this work is a promising initial step and proof-of-concept in developing a robust flare detection algorithm that does not rely on sigma-clipping or iterative approaches. There many avenues that we plan to explore in future work:

\begin{itemize}

    \item We aim to speed up the computation time so that we can apply our HMM approach to a large sample of M-dwarf stars observed by TESS and recover stellar flares.
    
    \item In a follow-up paper, we will estimate both the flare energy distribution and flare frequency distribution (FFD) through the posterior distribution of the Viterbi state-decoded sequences. Through this approach, we will be able to propagate the uncertainties in the duration of the flare from the state sequences to the energy of the flare in a coherent way. This, combined with the improved sensitivity and PPV of our method in detecting small and large flares, should produce better estimates and increase our confidence regarding the FFD of M-dwarf stars.

    \item Further work may explore incorporating multivariate data (i.e., time series data across multiple bands) through a hierarchical Bayesian paradigm.
    
    \item Ultimately, it could be fruitful to design a hierarchical model that includes the FFD parameters at the population level. In this way, many M-dwarf stars could be fit with our HMM approach simultaneously, and both their individual parameters and the population-level parameters of the FFD would be modeled in a coherent way.
    
    \item After detecting stellar flares in a star, and obtaining the state-decoded sequences that quantify the uncertainty in the duration of each flare for that star, one could still use stellar flare templates to model the flares and obtain parameter estimates of interest. A study comparing this approach with standard approaches in the literature could provide further insight into the benefits (or not) of using HMMs in this framework.

    \item The HMM technique presented here could also be further developed for time series data measured across multiple bands, which is now becoming more commonplace \citep[e.g.,]{joseph2024simultaneous}. This would allow a HMM analysis of stellar flares measured by future data sets \citep[e.g., CubeSat][]{poyatos2023observing}.
    
\end{itemize}

To our knowledge, this paper is one of the first applications of HMMs to an astronomy problem, and the first to do so for stellar flare detection. This statistical method has promise not only for stellar flare detection but also for other areas in astrophysics with time series data, such as gamma ray bursts, fast radio bursts, and quasars. Our hope is that this paper is useful as a starting-off point for the astronomical community to use this method in both stellar flare detection and other areas of astronomy.

\section*{Author Contributions}

J.A.E and Y.S. contributed equally to this project. Y.S. wrote the original code, performed most of the analysis, and wrote parts of the paper. J.A.E. subsequently modified the code, checked and performed additional analyses, wrote all code and analysis to produce all of the figures (except Figure~\ref{fig:graph}), and contributed substantially to writing Sections~\ref{sec:method}, \ref{sec:results}, and \ref{sec:discussion}. 

V.L.B. provided expertise on HMMs and co-supervised J.A.E. and Y.S.. V.L.B also wrote parts of Section~\ref{sec:HMMs} and made Figure~\ref{fig:graph}. V.L.B. was also a co-I on the DSI grant (see Acknowledgements) that funded this project.

G.M.E. was the PI on the DSI grant that funded this work. G.M.E. also co-supervised J.A.E. and Y.S., and provided expertise on Bayesian anaylsis and astrostatistics. G.M.E. edited and contributed to all sections of the paper, and contributed substantially to Section~\ref{sec:conclusion}.

J.S. co-supervised J.A.E. and Y.S.. J.S. contributed to the data handling and interpreation of results, and provided comments on the manuscript. 

R.V.C. was a co-I on the DSI grant that funded this work. R.V.C. provided expertise on Bayesian analysis and inference, and provided comments on the manuscript.

A.M. co-supervised Y.S., and contributed substantial expertise on M-dwarfs. A.M. also helped write parts of Section~\ref{sec:intro}.

J.R.A. D. provided expertise on M-dwarfs and stellar flares, provided scientific comments and feedback, and contributed to the editing of the final paper.

\FloatBarrier
\noindent \textbf{Acknowledgements.} 
Some computations involved in this paper were performed using the compute resources and assistance of the UW-Madison Center For High Throughput Computing (CHTC) in the Department of Computer Sciences. Computing resources in the Department of Statistical Sciences at the University of Toronto were also used, with support from IT specialist Claire Yu. This work was supported by the University of Toronto DSI catalyst fund to G.E. (PI), V.L.B., R.C., and A.M..

\bibliographystyle{aasjournal} 
\bibliography{references}

\end{document}